\documentclass[pra,aps,showpacs,twocolumn,floatfix]{revtex4}
\usepackage{amsmath,amssymb,amsfonts,graphicx}

\newcommand{\ac}{a^{\dag}}

\def\bmsigma{\boldsymbol{\sigma}}

\def\bmX{\boldsymbol{X}}
\def\bmR{\boldsymbol{R}}

\def\Tr{\hbox{Tr}}
\begin{document}
\title{Quantifying non-Markovianity of continuous variable Gaussian dynamical maps}
\author{Ruggero Vasile}\email{ruggero.vasile@utu.fi}
\affiliation{Turku Centre for Quantum Physics, Department of Physics
and Astronomy, University of Turku, FI-20014 Turun yliopisto,
Finland.}
\author{Sabrina Maniscalco}
\affiliation{Turku Centre for Quantum Physics, Department of Physics
and Astronomy, University of Turku, FI-20014 Turun yliopisto,
Finland}\affiliation{ School of Engineering \& Physical Sciences,
Heriot-Watt University Edinburgh, EH144AS, UK}
\author{Matteo G.~A.~Paris}
\affiliation{Dipartimento di Fisica, Universit\`a degli Studi di
Milano, I-20133 Milano, Italy}
\affiliation{CNISM, UdR Milano Statale, I-20133 Milano, Italy}
\author{Heinz-Peter Breuer}
\affiliation{Physikalisches Institut, Universit\"at Freiburg,
Hermann-Herder-Strasse 3, D-79104 Freiburg, Germany}
\author{Jyrki Piilo}
\affiliation{Turku Centre for Quantum Physics, Department of Physics
and Astronomy, University of Turku, FI-20014 Turun yliopisto,
Finland. }
\begin{abstract}
We introduce a non-Markovianity measure for continuous variable open
quantum systems based on the idea put forward in H.-P. Breuer {\em
et al.}~Phys.~Rev.~Lett.\textbf{103}, 210401 (2009), i.e., by
quantifying the flow of information from the environment back to the
open system. Instead of the trace distance we use here the fidelity
to assess distinguishability of quantum states. We employ our
measure to evaluate non-Markovianity of two paradigmatic Gaussian
channels: the purely damping channel and the quantum Brownian motion
channel with Ohmic environment. We consider different classes of
Gaussian states and look for pairs of states maximizing the backflow
of information. For coherent states we find simple analytical
solutions, whereas for squeezed states we provide both exact
numerical and approximate analytical solutions in the weak coupling
limit.
\end{abstract}
\date{\today}
\maketitle
\section{Introduction}
Physical systems are never perfectly isolated from their
environment. Especially in quantum mechanics, when they are
exploited to perform quantum computation or communication protocols,
the interaction of the systems of interest with the environment
should be considered in the derivation of the dynamical equations. Effects
of this interaction, e.g. decoherence and disentanglement, can
indeed endanger the accomplishment of any task based on quantum
features like coherence or entanglement.
\par
In order to take into account the presence of the environment and
its influence on the dynamics of the system, within the theory of
open quantum systems~\cite{BrePet,Weiss} a variety of techniques
have been developed to describe the evolution of the system of
interest, e.g.~by quantum master equations. The functional form of
any master equation depends both on the system and environment, and
on the specific features and strength of the interaction. In the
literature the dynamics of open quantum systems are often described
using master equations in the so-called Lindblad
form~\cite{LiGoKoSu}. Profitable applications of this class of
dynamical equations are present in many fields of physics, and
systems whose dynamics is described by equations in the Lindblad
form are generally called Markovian. As it will be explained in more
detail in Sec. II, the dynamical maps associated with Lindblad
master equations are divisible, implying that during the evolution
any pair of initially different states becomes less and less
distinguishable. This phenomenon is interpreted as an irreversible
loss of information which flows from the system to the environment,
and it is considered to be the key feature of
Markovianity~\cite{Bre09}.
\par
In practice, however, Lindblad master equations are derived under a
series of approximations. The exact dynamics of any physical system
is generally described by other classes of master equations.
Recently, a lot of effort has been devoted to provide a formal
definition of non-Markovianity in open quantum systems, e.g.~to
capture physical features such as the re-coherence due to reservoir
memory effects~\cite{NMQJ,pseudo}. These
efforts~\cite{Wolf08,Bre09,Rivas10} also lead to computable measures
for the degree of non-Markovianity. In this paper, we focus on the
definition given in Ref.~\cite{Bre09} where non-Markovianity is
defined in terms of the information flow between the open system and
its environment.
\par
Besides its own importance from a purely theoretical point of view,
the concept of non-Markovianity and its quantification may also find
practical applications. One may ask indeed, whether non-Markovianity
can be considered as a resource to improve quantum technologies.
More specifically, assuming that the density of modes of the
reservoir may be engineered in a controlled way to induce
non-Markovian behavior, can this be used to improve existing quantum
protocols? The first affirmative answers come from quantum metrology
and quantum key distributon. In Ref.~\cite{Chin11} the authors
investigate the problem of parameter estimation when the quantum
channel is non-Markovian according to the definition given in
Ref.~\cite{Rivas10}. They find that, for some non-Markovian
reservoirs, the estimation can be improved compared to the Markovian
case. The other example is reported in Ref.~\cite{Vas11} where it
has been proven that quantum key distribution protocols in
non-Markovian channels provide alternative ways of protecting the
communication which cannot be implemented in usual Markovian
channels.
\par
Even if the definition introduced in Ref.~\cite{Bre09} is
independent of the nature of the physical system, it has been
applied so far to the discrete variable case only. In this paper we
extend the analysis to continuous variable (CV)
systems~\cite{BraunRev} which, in quantum information and
communication, represent a valid, and sometimes better, alternative
to discrete variable systems.  Our aim is to introduce and study a
computable measure for the degree of non-Markovianity in continuous
variable systems focussing on some relevant examples of Gaussian
preserving maps. Moreover we also consider the possibility of
evaluating the map only for subsets of Gaussian states (e.g.
coherent states and squeezed states) with the main intent to provide
a characterization of the map for protocols relying only on those
specific classes of states. This approach paves the way to a
definition of non-Markovianity as a resource in quantum information
theory.
\par
The paper is organized as follows: in Sec.~II we review the
non-Markovianity measure we use and extend the definition to the
realm of continuous variable Gaussian states. In Sec.~III we focus
on a phenomenological master equation describing a damping channel
and evaluate its non-Markovianity, whereas in Sec.~IV we address the
same issue for quantum Brownian motion in the weak coupling limit.
Finally in Sec.~V we discuss the results and close the paper with
some concluding remarks.
\section{Quantifying non-Markovianity in continuous variable systems}
The measure for the degree of non-Markovianity of a quantum process
introduced in~\cite{Bre09} is based on the distinguishability of two
different initial quantum states $\rho_1$ and $\rho_2$ under the
action of the open system dynamical map $\Phi_t$ associated to the
process. The distinguishability is qualified and quantified
in~\cite{Bre09} through the introduction of a proper distance
measure between quantum states, the trace distance defined as
$\mathcal{D}(\rho_1,\rho_2)=\Tr|\rho_1-\rho_2|/2$. The trace
distance satisfies a contractivity property under the action of any
completely positive (CPT) map $\Phi$
\begin{equation}
\mathcal{D}(\rho_1,\rho_2)\geq \mathcal{D}(\Phi\rho_1,\Phi\rho_2).
\end{equation}
Given a dynamical map $\Phi_{t,t_0}$, this is called divisible if
the evolution up to a time $t$ can be written as a completely
positive evolution from the initial time $t_0$ to an intermediate
time $\tau$, and another completely positive evolution from the
intermediate time to the final time $t$, i.e.
$\Phi_{t,t_0}=\Phi_{t,\tau}\cdot\Phi_{\tau,t_0}$, for any
$t_0<\tau<t$. Lindblad dynamical semigroups \cite{LiGoKoSu} describe
divisible processes. It follows that under such dynamics the trace
distance is always monotonic, i.e.
$\mathcal{D}\bigl(\rho_1(\tau),\rho_2(\tau)\bigl)\geq
\mathcal{D}\bigl(\rho_1(t),\rho_2(t)\bigl)$ for any pair
$\rho_1(0)$, $\rho_2(0)$ of initial states and for any
$t_0\leq\tau\leq t$. The monotonic decrease however holds also for
classes of divisible maps which are not of a Lindblad form (examples
are given in~\cite{Bre09}). On the other hand, the most dynamical
evolutions violate both the divisibility condition and the
monotonicity property of the trace distance. When monotonicity is
not satisfied it means that there are intervals of time for which
the states become more distinguishable compared to previous
instants. This feature is interpreted as a flow of information from
the environment back to the system, a striking property which
characterizes a non-Markovian evolution. The measure of the degree
of non-Markovianity is then defined as
\begin{equation}\label{NMDEF}
\mathcal{N}=\max_{\rho_1,\rho_2}\int_{\dot{\mathcal{D}}>0}
\frac{d}{dt}\mathcal{D}(\rho_1(t),\rho_2(t))dt,
\end{equation}
where $\dot{\mathcal{D}}$ indicates the time derivative and the
maximization is taken over all the possible pairs of initial states.
\par
So far the quantity in \eqref{NMDEF} has been evaluated and analyzed
in some details for discrete variables quantum maps, e.g. a one
qubit channel~\cite{Bre09,Bas11,Znidaric11}. In this paper we extend
it to continuous variable systems, focusing to single-mode systems.
The extension involves two main issues requiring specific attention.
The first comes from the fact that the Hilbert space for continuous
variable systems is infinite dimensional, and therefore it is not
possible to characterize all the states with a finite number of
parameters as in the qubit case (e.g. Bloch sphere representation).
The second issue is related to the lack of an analytic expression
for the trace distance or other, equivalent, distance measures for a
generic CV state. On the other hand, these issues may be solved upon
restricting the analysis to Gaussian states, and Gaussian preserving
channels~\cite{Gauss}. In fact, Gaussian states can be uniquely
characterized by a finite number of parameters. Moreover, since
analytic expressions for the trace distance are lacking, alternative
distinguishability signatures may be employed within the same
spirit. One possible choice is to use the fidelity
\begin{equation}\label{Fidelity}
\mathcal{F}(\rho_1,\rho_2)=\Tr\sqrt{\sqrt{\rho_1}\rho_2\sqrt{\rho_1}},
\end{equation}
which is related to a proper distance measure, the Bures distance
$\mathcal{D}_{\mathcal{F}}(\rho_1,\rho_2)=
\sqrt{2-2\sqrt{\mathcal{F}(\rho_1,\rho_2)}}$. Remarkably, the
fidelity, and, thus, also the Bures distance, are monotonic under
the action of any CPT map, making it a good candidate for taking the
role of the trace distance in the definition of the non-Markovian
measure.
\par
An alternative choice we could consider is the relative entropy
$\mathcal{S}(\rho_1||\rho_2)=\Tr\bigl[\rho_1\log\rho_1-\rho_1\log\rho_2\bigl]$
whose expression for Gaussian states is known \cite{Mar04}. Despite
the fact that also the relative entropy possesses a contractivity
property under CPT maps, it is not a proper distance measure (e.g.
it lacks symmetry property), and also it is not bounded. For these
reasons we base our study on the  fidelity.
\par
The most general single-mode Gaussian state can be written as
\cite{Gauss}
$$\rho^G = D(\beta)S(\xi) \nu_{\rm
th}(N)S^{\dag}(\xi)D^{\dag}(\beta)\,,$$ where $S(\xi) = \exp[\frac12
(\xi{a^{\dag 2}}-\xi^*a^2)]$ and $D(\beta) = \exp[ \beta a^{\dag} -
\beta^*a)]$ are the squeezing operator and the displacement
operator, respectively, and $\nu_{\rm th}(N) = (N)^{a^\dag
a}/(1+N)^{a^\dag a + 1}$ is a thermal equilibrium state with $N$
average number of quanta, $a$ being the annihilation operator. Upon
introducing the vector operator $\bmR^T=(R_1,R_2)\equiv(q,p)$, where
$q=(a+a^\dag)/{\sqrt{2}}$ and $p=(a^\dag-a)/(i\sqrt{2})$ are the
so-called quadrature operators, we can fully characterize $\rho^G$
by means of the first moment vector $$\overline{\bmX}^T =
\langle\bmR^T \rangle = \sqrt{2}(\Re{\rm e}[\beta],\Im{\rm
m}[\beta])\,,$$ where $\langle A \rangle = \Tr[A\, \rho]$, and of
the $2\times 2$ covariance matrix (CM) $\bmsigma$, with elements
$$[\bmsigma]_{hk}=\frac12\langle R_hR_k + R_kR_h \rangle - \langle
R_h\rangle\langle R_k \rangle \quad k=1,2\,.$$
\par
The expression of the fidelity for a generic pair of Gaussian states
$\rho_1^G$, $\rho_2^G$ can be given in a closed analytical form \cite{Scu98}
and, as expected, depends only on the values of the vectors
$\overline{\bmX}_i$, and covariance matrices $\bmsigma_i$ of the
states involved
\begin{equation}\label{FideG}
\mathcal{F}(\rho^G_1,\rho^G_2)=\frac{2}{\sqrt{\Delta+\delta}-\sqrt{\delta}}
\, e^{-\frac{1}{2}\overline{\bmX}^T(\bmsigma_1+
\bmsigma_2)^{-1}\overline{\bmX}},
\end{equation}
where
\begin{align}
\Delta & =4\,\hbox{Det}(\bmsigma_1+\bmsigma_2), \\
\delta &
=16\left[\hbox{Det}(\bmsigma_1)-\frac14\right]\left[\hbox{Det}
(\bmsigma_2)-\frac14\right].
\end{align}
The fidelity in Eq. \eqref{FideG} is a function of all the
parameters that characterize the pair of Gaussian states: two
complex displacement amplitudes $\beta_i=|\beta_i|e^{i\theta_i}$,
two complex squeezing amplitudes $\xi_i=r_ie^{i\phi_i}$ and two real
thermal parameters $N_i$. In order to simplify the notation used in
the following we introduce here a set of collective arguments
\begin{equation}\begin{split}
&\mathbf{P_N}\equiv\{N_1,N_2\}\,,\\
&\mathbf{P_S}\equiv\{r_1,r_2,\phi_1,\phi_2\}\,,\\
&\mathbf{P_C}\equiv\{|\beta_1|,|\beta_2|,\theta_1,\theta_2\}\,,
\end{split}\end{equation}
and denote, e.g. by $\mathcal{F}(\{\mathbf{P_C},\mathbf{P_N}\},t)$
the fidelity between two mixed coherent states at time $t$ of the
evolution. The full set of parameters is denoted by the symbol
$\mathbf{P}$.
\par
A non-Markovianity measure may be obtained by integrating the time
derivative of the fidelity $\mathcal{F}(\mathbf{P},t)$ over the
intervals in which it decreases. If the class of initial states is
characterized by the set of parameters $\mathbf{P}$ the measure may
be written as
\begin{equation}\label{MeasGa}
\mathcal{N}_{\mathbf{P}}=\max_{\mathbf{P}}\,
\biggl[-\int_{\dot{\mathcal{F}}<0}\frac{d}{dt}
\mathbf{\mathcal{F}}(\mathbf{P},t)\,dt\biggl],
\end{equation}
where $\dot{\mathcal{F}}$ indicates the time derivative and the
maximization is taken over the set of parameters $\mathbf{P}$.
\par
The measure $\mathcal{N}_{\mathbf{P}}$ in \eqref{MeasGa} is obtained
by maximization over the class of Gaussian states, a procedure which
makes it particularly suitable to asses Gaussian preserving
channels. On the other hand, when applied to a generic channel, it
cannot be considered as a global property. This is not a crucial
issue for practical applications for at least two reasons. On the
one hand, this choice allows to actually calculate and compare the
degree of non-Markovianity for continuous variable channels, a task
that would not be feasible for non-Gaussian states. On the other
hand, it should be noticed that while in principle there are
techniques to prepare any kind of single qubit states, the same does
not apply to continuous variable systems. As a matter of fact the
class of Gaussian states plays a crucial role in quantum information
processing, since they can be characterized theoretically in a
convenient way, and they can also be generated and manipulated
experimentally in a variety of physical systems, ranging from light
fields to atomic ensembles. Therefore, our aim in the rest of the
paper will be that of characterizing the \emph{Gaussian degree of
non-Markovianity} for some relevant Gaussian preserving channels.
\par
Once we restrict the investigation to Gaussian states we still have
to face the problem of the maximization procedure, which may be
challenging from the numerical point of view, since the domains of
some of the involved parameters are unbounded. One way to deal with
this issue is to focus on subclasses of Gaussian states, e.g. pure
coherent states or squeezed states, and therefore reduce the number
of parameters involved. On the other hand, it is also possible to
bound the domain of definition of the parameters, invoking the same
line of reasoning used previously: experimental accessibility. For
example, in practice it is not possible to obtain an arbitrary
amount of squeezing~\cite{Ebe10}. The consequence is then a
limitation of the domain of definition and therefore a faster
convergence of numerical maximization algorithms. In the cases we
examine in the next sections however we will see that it is not
always needed to bound the domain of definition of the parameters.
This is because the maximizing pair of states depends on the
strength of the interaction, i.e., the coupling constant, and for
weakly coupled systems, experimentally accessible values for the
squeezing and displacement may characterize the maximizing pair.
\par
In the following Sections we will assume that the maximum is
achieved for pure states, i.e., we assume $N_1=N_2=0$ and perform
the maximization over the other parameters. This assumption may be
proved for the case of coherent thermal states in the weak coupling
regime, whereas we conjecture its validity for the other classes of
states.
\par
In the next two sections we introduce two different examples of
master equations used to describe the dynamics of continuous
variable systems and we study their non-Markovian Gaussian
properties.
\section{Damping master equation}
We start by considering the dynamics described by the following
phenomenological Lindblad type equation with a single decay channel
and a time dependent damping rate $\gamma(t)$,
\begin{equation}\label{DampME}
\frac{d\rho}{dt}=\alpha\frac{\gamma(t)}{2}(2a\rho\ac-\ac
a\rho-\rho\ac a)\,.
\end{equation}
Any Gaussian state evolving according  Eq. (\ref{DampME}) remains
Gaussian, with the displacement and the covariance matrix evolving
as follows  \cite{Gauss}
\begin{align} \label{EvoDamME}
\beta(t)  & =e^{-\frac{x(t)}{2}} \beta(0),\\
\sigma(t) & =e^{-x(t)} \sigma(0)+ \left[1-e^{-x(t)}\right]
\frac{\mathbb{I}}{2}, \notag
\end{align}
where
\begin{equation}
x(t)=\alpha\int_0^t2\gamma(s)ds,
\end{equation}
and $\alpha$ being a coupling constant. If $x(t)\ll1$ then we can
approximate $e^{-x(t)}\simeq 1-x(t)$. Under this weak coupling
condition we can also approximate the solution of \eqref{DampME} as
follows
\begin{equation}
\begin{split}
\label{EvoDamMEWC}
&\beta(t)=\biggl[1-\frac{x(t)}{2}\biggl]\beta(0),\\
&\sigma(t)=\bigl[1-x(t)\bigl]\sigma(0)+x(t)\,\frac{\mathbb{I}}{2}\,.
\end{split}
\end{equation}
Upon inserting Eq.~\eqref{EvoDamMEWC} into Eq.~\eqref{FideG} we
obtain the expression for the fidelity in the weak coupling limit.
The divisibility property of \eqref{DampME} here is equivalent to
the condition $\gamma(t)\geq0$ for any $t\geq0$, as it can be easily
verified from the solution \eqref{EvoDamME}. On the other hand, the
condition for non-Markovianity can be studied by inspecting the
derivative of the fidelity with respect to time. Because the
evolution of any Gaussian state depends on time through the function
$x(t)$ only, we have
\begin{equation}\label{DerFidDam}
\frac{d\mathcal{F}}{dt}=\frac{d\mathcal{F}}{dx}
\frac{dx}{dt}=2\alpha\gamma(t)\frac{d\mathcal{F}}{dx}.
\end{equation}
If $\gamma(t)=\gamma_0>0$ and therefore $x_M(t)=2\alpha\gamma_0 t$,
then expression \eqref{DerFidDam} must always be positive, because
the dynamics is described by a divisible map. Therefore for any
$t>0$ and any initial pair of states we must have
$\frac{d\mathcal{F}}{dx}>0$. Because the image sets of $x(t)$ and
$x_M(t)$ are the same, the condition $\frac{d\mathcal{F}}{dx}>0$
must hold also in the case of a time dependent decay rate. Stated in
another way, Eq. \eqref{DampME} describes a non-Markovian channel if
and only if the corresponding dynamical map is non-divisible. This
condition is valid for any pair of initial states. The amount of
non-Markovianity of the channel, as defined by Eq. \eqref{MeasGa},
can be written as
\begin{equation}\label{MeasDa}
\mathcal{N}_{\mathbf{P}}=\max_{\mathbf{P}}\sum_{I}\bigl[\mathcal{F}
(\mathbf{P},t^+_I)-\mathcal{F}(\mathbf{P},t^-_I)\bigl],
\end{equation}
where $[t^+_I,t^-_I]$ is the $I$-th negativity interval of
$\gamma(t)$.
\par
For coherent states we may derive an exact expression for
non-Markovianity valid for any form of the damping rate. Assuming a
single interval of negativity, we have
\begin{equation}\begin{split}\label{ExNMCoh}
&\mathcal{N}_{\mathbf{P_C}}=\exp\bigl\{-K
e^{-x(t^+)}\bigl\}-\exp\bigl\{-K e^{-x(t^-)}\bigl\},\\
&K=\frac{|\beta_1e^{i\theta_1}-\beta_2|^2}{2}=
\frac{x(t^-)-x(t^+)}{e^{-x(t^+)}-e^{-x(t^-)}}\,,
\end{split}\end{equation}
where only one angle appears in the parameter $K$ because the master
equation is invariant under a rotation in phase space. Other
analytic results can be also obtained if we consider small values of
the coupling constant. For example in the case of coherent states
and for any number of negativity periods of the damping rate, we
have
\begin{equation}\begin{split}\label{DamCohMeas1}
\mathcal{N}_{\mathbf{P_C}}&\simeq\max_{\mathbf{P}_C}
\sum_{I}\bigl[e^{-[1-x(t^+_I)]K}-e^{-[1-x(t^-_I)]K}\bigl]\\
&= \max_{\mathbf{P}_C}e^{-K}\sum_{I}\bigl[e^{x(t^+_I)K}-e^{x(t^-_I)K}\bigl]\\
&=\alpha\,\biggl\{\max_{\mathbf{P_C}}
f_1(\mathbf{P_C})\biggl\}\,\int_{\gamma<0}\!\!2\gamma(t)\,dt+o(\alpha^2),
\end{split}\end{equation}
where we define the state dependent function
$f_1(\mathbf{P_C})=e^{-K}K$. From Eq.$\,$\eqref{DamCohMeas1} we can
conclude that to first order in the coupling $\alpha$, the quantity
$\mathcal{N}_{\mathbf{P}_C}$ is proportional to the integral of the
negativity region of the damping rate, and the next contribution
appears only to third order. The measure is then maximized for
$K=1$, a condition which defines a whole set of pairs of states, all
of them leading to the same value of $\mathcal{N}_{\mathbf{P}_C}$ to
first order in $\alpha$. We then conclude that the degree of
non-Markovianity for the set of coherent states can be written as
\begin{equation}\begin{split}\label{DamCohMeas2}
\mathcal{N}_{\mathbf{P_C}}=\frac{2}{e}\alpha
\int_{\gamma<0}\!\!\gamma(t)\,dt+o(\alpha^2)\,.
\end{split}\end{equation}
This result has actually been derived under the hypothesis that also
$e^{x(t^{\pm}_I)K}\simeq 1+x(t^{\pm}_I)K$ for any interval $I$,
therefore limiting the parameter domain depending on the value of
$\gamma(t)$ and $\alpha$. However, in the weak coupling limit the
domain is large enough to contain the points for which $K=1$.
\par
We now turn to the class of pure squeezed vacuum states (from now on
referred to as squeezed states), characterized by the set of
parameters $\mathbf{P_S}\equiv\{r_1,r_2,\phi_1,\phi_2\}$, which, in
the absence of displacement, can be reduced to
$\mathbf{P_S}\equiv\{r_1,r_2,\phi,0\}$, with $\phi$ the angle
between the squeezing directions. We can follow the same line of
reasoning used previously for coherent states and, e.g. for a single
period of negativity of the damping coefficient $[t^+,t^-]$, define
the quantity
\begin{equation}\begin{split}\label{DamSqueMeas1}
\mathcal{N}_{\mathbf{P_S}}&=\max_{\mathbf{P_S}}
\bigl[\mathcal{F}(\mathbf{P_S},t^+)-\mathcal{F}(\mathbf{P_S},t^-)\bigl]\\
& =\alpha\, \biggl\{\max_{\mathbf{P_S}}g_1(\mathbf{P_S})\biggl\}
\int_{\gamma<0}\!\!2\gamma(t)\,dt+o(\alpha)\, ,
\end{split}\end{equation}
where function $g_1(\mathbf{P_S})$ depends on the parameters
$r_1,r_2$ and $\phi$ in a rather complicated way. Its expression can
be simplified if we set $r=r_1=r_2$
\begin{equation}\label{G1PS}
g_1(\mathbf{P_S})=8\cosh(2r)\frac{k(r,\phi)-\sqrt{k(r,\phi)}}{k^2(r,\phi)},
\end{equation}
where $k(r,\phi)=3+\cos\phi+\cosh4r(1-\cos\phi)$. We will discuss
later on in this Section about the maximization of the function
$g_1(\mathbf{P_S})$.

From Eq. \eqref{DamSqueMeas1} we notice that the second order term
is not zero for squeezed states, suggesting that the first order
approximation has a more limited validity than in the case of
coherent states. It is also nontrivial to determine the domain of
parameters which allows the expansion to be truncated at first
order. This is due to the more complicated functional form of the
fidelity when squeezing, instead of displacement is implemented. The
first order result is then of a similar form to the coherent state
case, with a different state dependent coefficient
$g_1(\mathbf{P_S})$. It is straightforward to show that this result
is independent of the number of negativity periods of $\gamma(t)$.
\par
Interesting results can be also derived if we consider now the most
general pure Gaussian state (both displacement and squeezing). From
\eqref{FideG} we can separate the fidelity as a product of two parts
$\mathcal{F}(\{\mathbf{P_C},\mathbf{P_S}\},t)=C(\{\mathbf{P_C},
\mathbf{P_S}\},t)S(\mathbf{P_S},t)$,
where $C(\{\mathbf{P_C},\mathbf{P_S}\},t)$ is the exponential part,
containing the coherent state amplitudes, and $S(\mathbf{P_S},t)$ is
the fidelity for zero displacement. A first order expansion in
$\alpha$ shows that
\begin{equation}\begin{split}
&\mathcal{N}_{\{\mathbf{P_C},\mathbf{P_S}\}}=\alpha
\int_{\gamma<0}\!\!2\gamma(t)\,dt\\
&\times\max_{\{\mathbf{P_C},\mathbf{P_S}\}}
\bigl\{S(\mathbf{P_S},0)f_1(\mathbf{P_C})+
C(\{\mathbf{P_C},\mathbf{P_S}\},0)g_1(\mathbf{P_S})\bigl\}\\
&+o(\alpha).
\end{split}\end{equation}
This result gives the non-Markovianity measure at first order in
$\alpha$ with a pair dependent coefficient given by a linear
combination of the displacement contributions $f_1(\mathbf{P_C})$
and the squeezing one $g_1(\mathbf{P_S})$. However, we cannot
conclude that even at this order the two contribution are completely
independent, because the weights appearing in the combination depend
on the fact that we applied squeezing and displacement. For example
it is easy to show that
$S(\mathbf{P_S},0)=\mathcal{F}(\mathbf{P_S},0)$ is the initial
fidelity of the same pair of states with zero displacement, and
$C(\{\mathbf{P_C},
\mathbf{P_S}\},t)=\mathcal{F}(\{\mathbf{P_C},\mathbf{P_S}\},0)
/\mathcal{F}(\mathbf{P_S},0)$, is the ratio between the initial
fidelity and the initial fidelity without displacement.
\par
It is worth noticing that for coherent-thermal states with equal
thermal parameters, i.e.  $N_1=N_2=N$, at first order in the
coupling we obtain
\begin{equation}\begin{split}\label{DamSqueMeas2}
\mathcal{N}_{\{\mathbf{P_N},\mathbf{P_C}\}}&=
\max_{N\geq0}\frac{\mathcal{N}_{\mathbf{P_C}}}{2N+1}+o(\alpha^2),
\end{split}\end{equation}
which is maximized by pure states ($N=0$), thus supporting our
choice to restrict the analysis to pure states only.
\par
As we mentioned before many features of non-Markovianity for the
damping channels do not depend on the specific form of the damping
rate. However, for the sake of concreteness let us now consider an
example of the damping rate
\begin{equation}\label{Decayrateg}\begin{split}
&\gamma(t)=\frac{1}{2}\biggl\{
             \begin{array}{c}
               e^{-t/10}\sin t, \\
               e^{-\pi/4}, \\
             \end{array}\qquad\begin{array}{c}
                               {\rm if} \\
                                {\rm if}
                              \end{array}
             \qquad\begin{array}{c}
                                      t<5\pi/2, \\
                                      t\geq 5\pi/2,
                                    \end{array}
\end{split}\end{equation}
which is characterized by only one interval of
negativity, $[\pi,2\pi]$, and where we are definitely
in the weak coupling regime if $\alpha \lesssim 0.1$.
\par
\begin{figure}[h!]
\includegraphics[width=0.95\columnwidth]{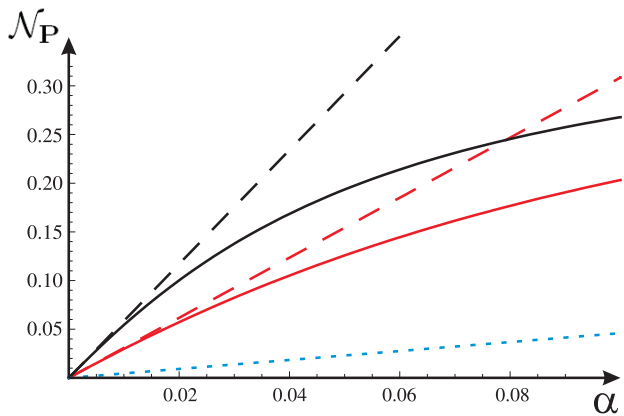}
\caption{(Color online) Non-Markovianity of the purely damping
channel as a function of the coupling $\alpha$. The dotted blue line
is the result obtained by maximizing over coherent states (the exact
solution and the first order approximation coincides). The solid
lines are the exact non-Markovianities for squezeed states, whereas
the dashed lines are the corresponding first order approximations.
Black curves correspond to $\phi=0.2$,while red ones to $\phi=0.1$.
\label{f:f1damp}}
\end{figure}

We start the analysis evaluating the non-Markovianity of the channel
for coherent states and squeezed states. These results are shown in
Fig.\ref{f:f1damp}, where we plot for both classes the
non-Markovianity evaluated at first order, together with the exact
numerical solution obtained by maximizing \eqref{MeasDa} using the
full solution \eqref{EvoDamME}. 
The coherent states maximization has been carried out over all the
parameters involved and their domain, while for squeezed states we
consider two fixed values of the angle $\phi$ between the squeezing
directions and then maximize over the magnitudes $r_1$ and $r_2$,
with the maximum value reached for $r_1=r_2$. Two main results are
evident: the non-Markovianity of the damping channel is larger for
squeezed states than for coherent states independently of the value
of coupling and increases with decreasing $\phi$. Due to the nature
of the master equation we can also state that this result is valid
independently of the form of $\gamma(t)$. The second result is that
the first order approximation is well sustained by coherent states
while it is violated by squeezed states also for relatively small
values of the coupling. This behavior for squeezed states becomes
more and more evident if we reduce the value of the angle $\phi$,
and it can be explained by inspecting Eq. \eqref{G1PS} for fixed
$\phi>0$. For small $\phi$, in fact, the function
$g_1(\mathbf{P_S})$ decreases, thus determining a violation of the
first order approximation. A further discussion of the results is
postponed to the final Section, after we have provided a more
complete picture by discussing the quantum Brownian motion model in
the next Section.
\section{Quantum Brownian motion}
The master equation describing quantum Brownian motion (QBM) in the
interaction picture, under the weak coupling and secular
approximations (see, e.g.~Ref.~\cite{QBMEq} and references therein),
is given by
\begin{equation}\begin{split}\label{QbmME}
\frac{d\rho}{dt}&=\alpha\,\frac{\Delta(t)+\gamma(t)}{2}(2a\rho\ac-\ac
a\rho-\rho\ac a)\\
&+\alpha\,\frac{\Delta(t)-\gamma(t)}{2}(2\ac\rho a-a \ac\rho-\rho
a\ac).
\end{split}\end{equation}
The diffusion and damping coefficients $\Delta(t)$ and $\gamma(t)$
can be derived once we provide the analytic form of the spectral
density $J(\omega)$ and the temperature $T$ of the environment,
assumed to be in a thermal state. Their expressions in the weak
coupling limit are
\begin{equation}\begin{split}\label{CoeffME1}
&\Delta(t)=\int_0^{t}\!\! ds\int_{0}^{\infty} \!\!\!
d\omega\, J(\omega)\biggl(N(\omega)+\frac{1}{2}\biggl)\cos(\omega_0s)\cos(\omega s), \\
&\gamma(t)=\int_0^{t}\!\! ds\int_{0}^{\infty}\!\! d\omega\,
J(\omega)\sin(\omega_0s)\sin(\omega s),
\end{split}\end{equation}
where $N(\omega)=(\exp\{\hbar\omega/k_BT\}-1)^{-1}$ is the mean
number of thermal photons for a mode of frequency $\omega$, and
$\omega_0$ is the bare frequency of the system. The spectral
function we are going to consider is the following Ohmic spectral
density
\begin{equation}\label{Spectrum}
J(\omega)\propto\omega\, C(\omega,\omega_c)\,,
\end{equation}
where $C(\omega,\omega_c)$ is a high frequency cutoff function with
$\omega_c$ being the cutoff frequency. Usually the cutoff function
is chosen to be of Lorentzian or exponential form. Here we use an
exponential cutoff $C(\omega,\omega_c)=\exp\{-\omega/\omega_c\}$,
such that
\begin{equation}
\label{ExpSpectrum}
J(\omega)=\omega e^{-\omega/\omega_c}\,.
\end{equation}
The exact solution of Eq. \eqref{QbmME} for the displacement and
covariance matrix of any initial Gaussian state is
\begin{equation}\begin{split}\label{EvoQbmEx}
&\beta(t)=e^{-\frac12 x(t)}\beta(0),\\
&\sigma(t)=e^{-x(t)}\sigma(0)+ \alpha\,
\frac{\mathbb{I}}{2}e^{-x(t)}\int_0^te^{x(s)}\Delta(s)ds\,,
\end{split}
\end{equation}
which, at first order in $\alpha$ can be written as
\begin{equation}\begin{split}\label{EvoQbmME}
&\beta(t)=\biggl[1-\frac{x(t)}{2}\biggl]\beta(0)\\
&\sigma(t)=\bigl[1-x(t)\bigl]\sigma(0)+y(t)\frac{\mathbb{I}}{2},
\end{split}\end{equation}
with
$$x(t)=2\alpha\int_0^t\!\!\gamma(s)\,ds, \quad
y(t)=2\alpha\int_0^t\!\!\Delta(s)\,ds\,,$$ and where
$\sigma(0)$ and $\beta(0)$ are the covariance matrix
and displacement vector of the initial state, respectively.
\par
The dynamics generated by Eq.~\eqref{QbmME} is more involved
compared to that of Eq.~\eqref{DampME}, and this gives us the
opportunity to study in more detail the non-Markovianity properties
of continuous variable systems. What is here relevant, compared to
the previous case, is the presence of two decay channels, one
downward channel and one upward. This structure leads to the
inequivalence of non-divisibility and non-Markovianity. For the
master equation~\eqref{QbmME} the divisibility property is satisfied
if $\Delta(t)\geq|\gamma(t)|$~\cite{Bre09}. On the other hand, we
have a non-Markovian behavior if the following quantity attains
negative values
\begin{equation}\label{NMQBM}
\frac{d\mathcal{F}}{dt}=\frac{\partial \mathcal{F}}{\partial
x}\frac{dx}{dt}+\frac{\partial \mathcal{F}}{\partial
y}\frac{dy}{dt}= 2\alpha\gamma(t)\frac{\partial
\mathcal{F}}{\partial x}+2\alpha\Delta(t)\frac{\partial
\mathcal{F}}{\partial y}.
\end{equation}
The sign of \eqref{NMQBM} depends in a non trivial way on the values
of the damping and diffusion coefficients, and of the derivatives of
the fidelity with respect to $x(t)$, $y(t)$, which are in general
functions of all the parameters and the time. Expression
\eqref{NMQBM} indicates that both damping and diffusion phenomena
contribute to the process. The dominance between the two
contributions depends on the spectral function $J(\omega)$, the
coupling constant, the temperature of the environment and on the
initial pair of states through the derivatives $\partial
\mathcal{F}/\partial x$ and $\partial \mathcal{F}/\partial y$. Each
derivative is proportional to the difference between two fidelities,
the one calculated for an increment in one variable and the initial
one. We expect that, e.g. $\mathcal{F}(x+h,y)>\mathcal{F}(x,y)$
holds,
because, according to Eq. \eqref{EvoQbmME}, the first term is the
fidelity between two states which have lost information about their
initial preparation [see Eq. \eqref{EvoQbmME}]. In other words, we
expect non-Markovianity to be dependent on the sign of the master
equation coefficients and only on the magnitude of the partial
derivatives.
\par
Numerical evaluation of the region of negativity of
$d\mathcal{F}/dt$ shows that the zeroes of the derivative are
essentially the same as those of the diffusion coefficient
$\Delta(t)$, thus suggesting that diffusion is the leading
phenomenon for non-Markovianity of QBM. In order to prove this
result we should inspect the form of the non-Markovianity at the
first order in the coupling $\alpha$, in the same spirit as in Sec.
III. For coherent states we have
\begin{equation}\begin{split}\label{QBMCohMeas}
\mathcal{N}_{\mathbf{P}_{\mathbf{C}}}=
\frac{2}{e}\alpha\int_{\Delta<0}\!\!\Delta(t)\,dt+o(\alpha^2),
\end{split}\end{equation}
and the contribution from the damping coefficient appears only in
the third power of the coupling. For squeezed states we get
\begin{equation}\begin{split}\label{QBMSqueMeas}
\frac{d\mathcal{F}}{dt}&= \alpha
S_{\Delta}(r_1,r_2,\phi)\Delta(t)+\alpha
S_{\gamma}(r_1,r_2,\phi)\gamma(t)+o(\alpha^2),
\end{split}\end{equation}
where the coefficients $S_{\gamma}(r_1,r_2,\phi)$ and
$S_{\Delta}(r_1,r_2,\phi)$ are the zeroth order expansions of the
derivatives $d\mathcal{F}/dx$ and $d\mathcal{F}/dy$, respectively.
An inspection on the magnitudes of the coefficients shows, however,
that unless $\phi\simeq 1$ (the maximum of the measure is instead
obtained in the region $\phi\ll1$), we have
$S_{\gamma}(r_1,r_2,\phi)\ll S_{\Delta}(r_1,r_2,\phi)$. Therefore,
non-Markovianity can be approximated by
\begin{equation}\begin{split}\label{QBMSqueMeas2}
    \mathcal{N}_{\mathbf{P}_{\mathbf{S}}}=\alpha\max_{r_1,r_2,\phi}
S_{\Delta}(r_1,r_2,\phi)
\int_{\Delta<0}\!\!2\Delta(t)\,dt+o(\alpha^2).
\end{split}\end{equation}
Within the validity of the first order expansion this explains why
the zeros of \eqref{NMQBM} essentially coincides with those of
$\Delta(t)$.
\par
As for the damping channel, it is possible to derive a closed
formula for the non-Markovianity of QBM for coherent states. The
only assumption is that the negativity of the derivative of the
fidelity coincides with the negativity of the diffusion coefficient.
For a single interval of negativity we have
\begin{align}
\mathcal{N}_{\mathbf{P_C}}=
&\exp\left\{-\frac{Pe^{-x(t^+)}}{e^{-x(t^+)}+y(t^+)}\right\}  \notag \\
-&\exp\left\{-\frac{Pe^{-x(t^-)}}{e^{-x(t^-)}+y(t^-)}\right\},
\end{align}
where
\begin{align}
P&=\frac{|\beta_1e^{i\theta_1}-\beta_2|^2}{2}=\ln\biggl[
\frac{e^{-x(t^+)}}{e^{-x(t^-)}}\frac{e^{-x(t^-)}+y(t^-)}
{e^{-x(t^+)}+y(t^+)}\biggl]\\\notag
&\times\biggl(\frac{e^{-x(t^+)}}{e^{-x(t^+)}+y(t^+)}-
\frac{e^{-x(t^-)}}{e^{-x(t^-)}+y(t^-)}\biggl)^{-1}\notag.
\end{align}
Because the non-Markovianity is here led by the diffusion phenomena,
we expect a strong dependence also on the temperature of the
environment. In the following, we will examine the non-Markovianity
in different temperature regimes, comparing the first order
approximation and the numerical one based on the exact solution
\eqref{EvoQbmEx}.
\par
The temperature dependence of the diffusion coefficients is apparent
in its definition \eqref{CoeffME}. Essentially the coefficient is a
sum of a zero temperature term and a contribution from the thermal
photons in the bath. Respectively, their expressions are
\begin{equation}\begin{split}\label{CoeffME}
&\Delta_0(t)=\frac{1}{2}\int_0^{t} ds\int_{0}^{\infty}
d\omega J(\omega)\cos(\omega_0s)\cos(\omega s),\\
&\Delta_T(t)=\int_0^{t} ds\int_{0}^{\infty} d\omega
J(\omega)N(\omega)\cos(\omega_0s)\cos(\omega s).
\end{split}\end{equation}
\begin{figure}[h!]
\includegraphics[width=0.8\columnwidth]{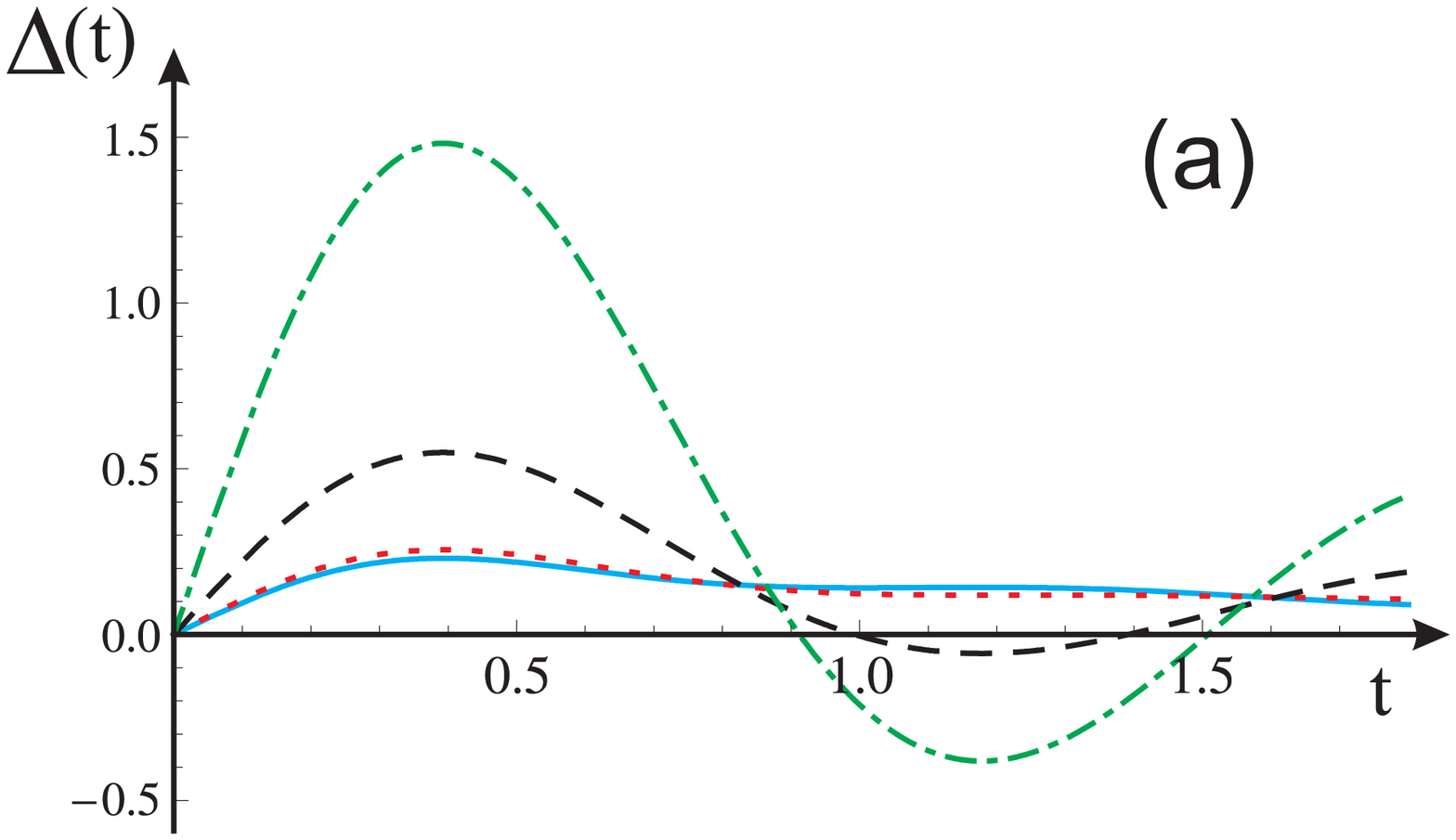}
\includegraphics[width=0.8\columnwidth]{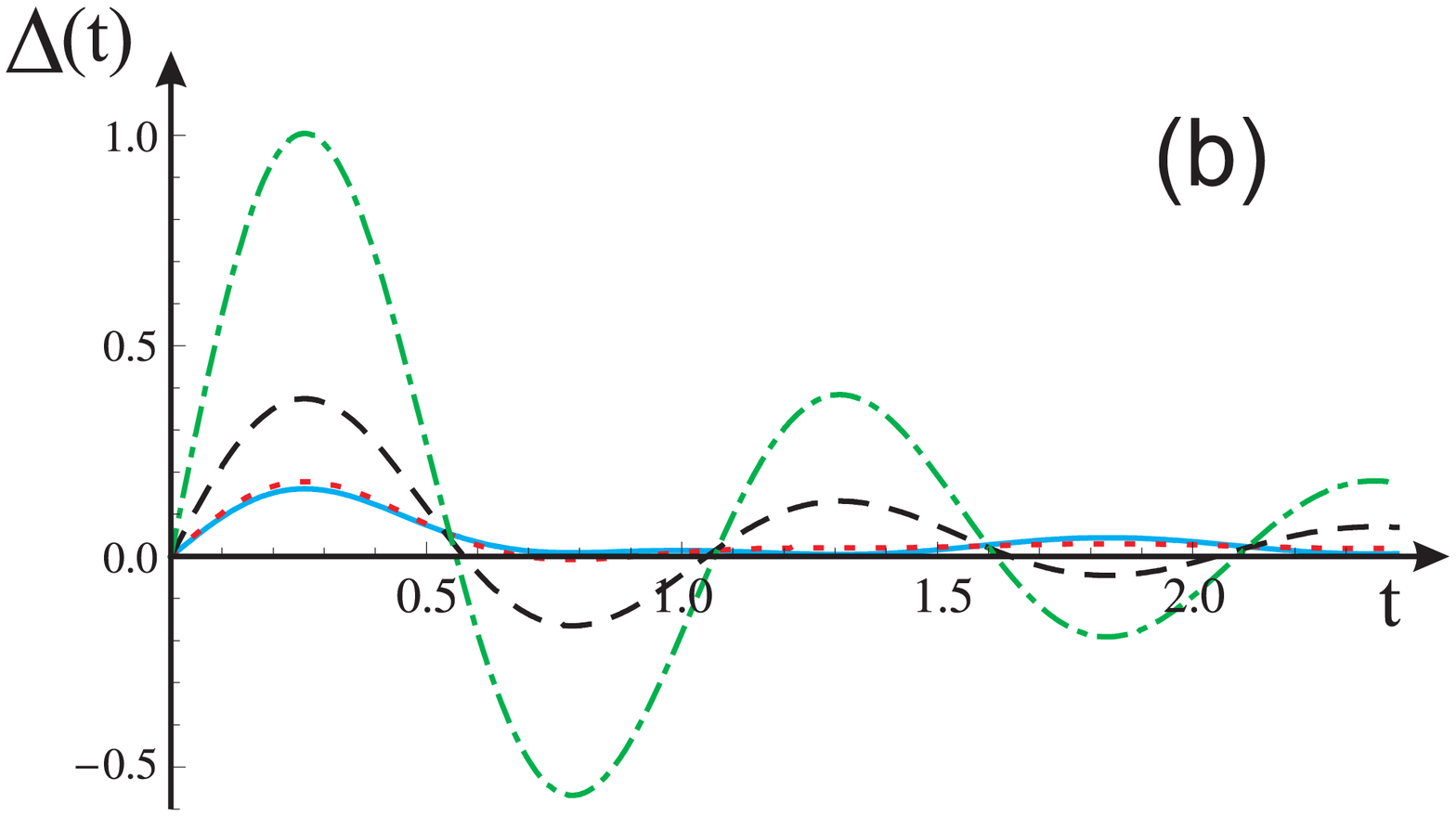}
\caption{(Color online) Diffusion coefficients as a function of time
for an Ohmic environment with $\omega_c=1$ and (a): $\omega_0=4$,
(b): $\omega_0=6$. We plot the zero temperature diffusion
coefficient (solid blue), low temperature ($k_BT/\hbar\omega_c=1/5$,
red dotted), intermediate temperature ($k_BT/\hbar\omega_c=1$, black
dashed) and high temperature ($k_BT/\hbar\omega_c=4$, green
dotted-dashed line).\label{Fig2}}
\end{figure} \\
For low temperature, that is when the thermal energy $k_BT$ is much
smaller than any excitation energy, $\hbar\omega_0,\hbar\omega_c$,
the total coefficient does not differ much from the zero temperature
contribution. Therefore we expect non-Markovianity to be independent
of $T$ in this regime. As $T$ is increased $\Delta_T(t)$ starts to
become relevant and in the high temperature regime becomes dominant
compared to $\Delta_0(t)$. In this regime the coefficient depends
linearly on $T$ and therefore we can expect a strong dependence of
non-Markovianity on temperature. Another feature is illustrated in
Fig.~\ref{Fig2}, where we show the diffusion coefficients for
different values of the bare frequency $\omega_0$. By comparing the
two panels, we can see that if we approach the resonance condition
$\omega_c\simeq\omega_0$, the diffusion coefficient at low
temperature does not show negative regions and so the first order
non-Markovianity is vanishing. As we increase the temperature,
$\Delta(t)$ may become negative and therefore the system shows a
non-Markovian behavior. This distinction between low and high
temperature ceases to be valid when we are out of resonance
$\omega_c/\omega_0\ll1$, where also for low temperature we can have
a non-Markovian behavior.
\begin{figure}[h!]
\includegraphics[width=0.9\columnwidth]{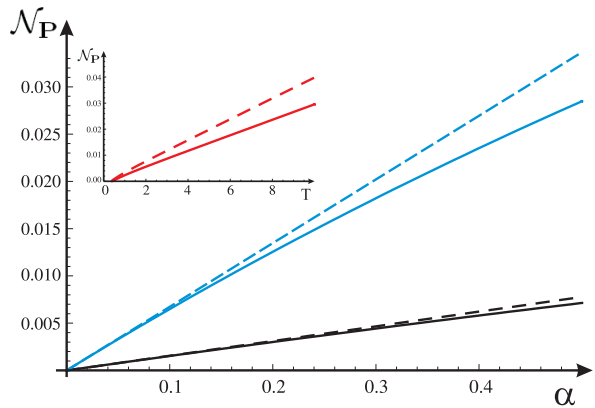}
\caption{(Color online) Non Markovianity of QBM channel with Ohmic
reservoir on coherent states as a function of the coupling $\alpha$
with $\omega_c=0.2$ and $\omega_0=1$, for different temperatures.
Dashed (solid) lines represent the first order approximate
(numerical) solutions. Black curves correspond to
$k_BT/\hbar\omega_0=0.2$ and blue curves to
$k_BT/\hbar\omega_0=0.5$. The inset shows the temperature dependence
for fixed coupling $\alpha=0.1$, and $\omega_0=1$ (red solid),
$\omega_0=1.1$ (red dashed).\label{Fig3}}
\end{figure}\\
After this discussion on the nature and behavior of the diffusion
coefficient, we are ready to illustrate the results about
non-Markovianity of the QBM channel. We start by considering the
class of coherent states, for which the first order measure is
defined in Eq.\eqref{QBMCohMeas}. In Fig.~\ref{Fig3} we show the
comparison between the analytic and numerical results for
$\omega_c=0.1$ and $\omega_0=1$, and for two different values of
temperature. When $\alpha<0.1$ the agreement is good in both cases,
whereas the first order approximation fails for the higher
temperature when we increase the coupling.  In the inset of
Fig.~\ref{Fig3} we show the first order measure for fixed coupling
($\alpha=0.1$) as a function of the temperature when $\omega_c=0.3$
and different values of $\omega_0$. When $T<0.4$ the measure is zero
indicating that the diffusion coefficient has no negativity regions.
\par
\begin{figure}[h!]
\includegraphics[width=0.9\columnwidth]{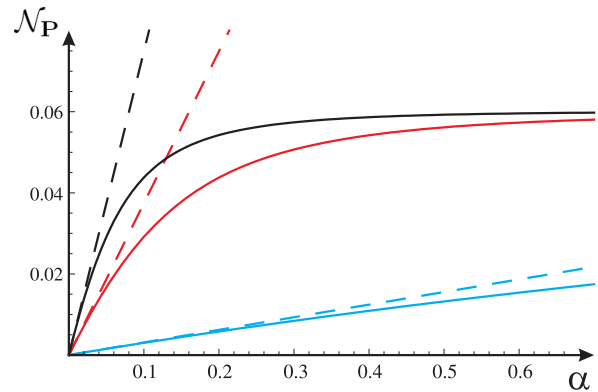}
\caption{(Color online) Non-Markovianity of QBM channel with Ohmic
reservoir as a function of $\alpha$. We have set
$k_BT/\hbar\omega_0=0.2$, $\omega_c=0.2$, $\omega_0=1$. Red and
black curves show the behavior of squeezed states, blue curves are
for coherent states. Dashed (solid) curves represent the first order
(numerical) solution to the maximization problem. For squeezed
states the maximization has been performed with the constraint
$r_1=r_2$ and for fixed phase $\phi=0.1$ (red curves), or
$\phi=0.05$ (black curves). For coherent states the maximization has
been performed over all their characterizing parameters.
\label{Fig4}}
\end{figure}

For squeezed states, in Fig. \ref{Fig4}  we show the comparison
between the numerical and first order analytic results for the
measure as a function of the coupling constant $\alpha$ for small
values of the temperature. We can notice a behavior similar to that
of the damping channel case (See Fig. 2a), the exact and
approximated expressions indeed coincide only for very small value
of the coupling.
\\
For increasing $\alpha$ the non-Markovianity saturates to a constant
value, which is achieved for smaller $\alpha$ when $\phi$ decreases.
In Fig. \ref{Fig4}  we compare the results for coherent and squeezed
states, showing that also for the QBM channel squeezed states are
more sensitive than coherent states to the non-Markovianity of the
channel.
\begin{figure}[h!]
\includegraphics[width=0.9\columnwidth]{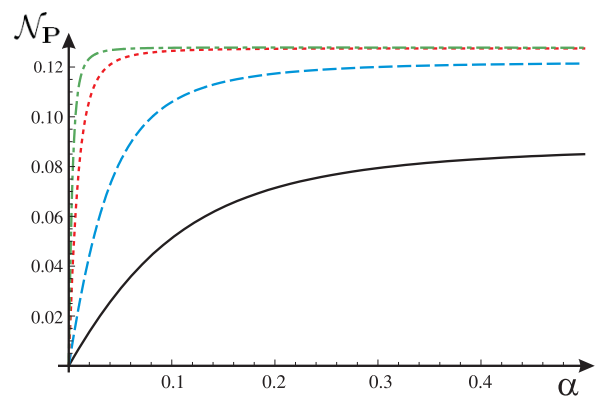}
\caption{(Color online) Non Markovianity of QBM channel with Ohmic
reservoir on coherent states. We have set $\omega_c=0.2$,
$\omega_0=1$ and maximized at fixed angle $\phi=0.05$ over the
squeezing parameters. Different lines correspond to different
temperatures: $k_BT/\hbar\omega_0=0.3$ (black solid),
$k_BT/\hbar\omega_0=0.9$ (blue dashed), $k_BT/\hbar\omega_0=4$ (red
dotted) and $k_BT/\hbar\omega_0=8$ (green dotted
dashed).\label{Fig5}}
\end{figure} \\
Finally in Fig. \ref{Fig5} we plot for fixed parameters
$\omega_c=0.2$, $\omega_0=1$ and $\phi=0.05$ the numerical results
for the measure for different temperatures. The behavior is
qualitatively the same, i.e. we have saturation for increasing
$\alpha$ and the saturation value increases with temperature, until
it reaches a maximum saturation value for high temperatures.

\section{Discussion and conclusions}
\label{s:out} In the previous Sections we have analyzed in detail
the non-Markovianity of two kinds of CV quantum channels, addressing
separately the non-Markovian behavior for coherent and squeezed
states. The dynamics of coherent states is governed by the evolution
of only the displacement amplitude in the damping channels, and by
both the displacement and the covariance matrix in the QBM channel
case. On the the other hand, the dynamics of (zero amplitude)
squeezed states depends on the covariance matrix only in both cases.
This behavior allows us to address the effects of the two
contributions on the amount of non-Markovianity of the channel.
\par
Despite the difference in the form of the master equations, many
common features are apparent. At first, we notice that squeezed
states are in general more sensitive to the non-Markovian behavior,
as witnessed by the larger non-Markovianity compared to coherent
states, independently on the value of the coupling and temperature.
Generally, for both classes of states, the non-Markovianity is
mostly related to the sign of the master equation coefficients. For
the case of QBM non-Markovianity is mostly due to diffusion
described by the coefficient $\Delta(t)$. On the other hand, even if
the dynamics of the displacement vector is not affected by it [see
Eq. \eqref{EvoQbmEx}], it is still fundamental for the class of
coherent states. This behavior is even more evident as the
temperature is increased, since the damping is independent of
temperature.
\par
Another interesting feature is the behavior of the non-Markovianity
for squeezed states as a function of the coupling and the
temperature, in particular for what concerns the quantum Brownian
motion case. As it is apparent from Fig. \ref{Fig5}, the value of
non-Markovianity as a function of the coupling saturates, with a
saturation value that increases with the temperature. This behavior,
whose origin may be traced back to the time evolution of the
covariance matrices, implies the existence of some bound on the flow
of information from the environment back to the open system as a
result of the Gaussian structure of the map.
\par
In conclusion, we have introduced a measure to quantify the
non-Markovianity of continuous variable quantum channels and
have used it to analyze two paradigmatic Gaussian channels: the
purely damping channel and the quantum Brownian motion channel with
Ohmic environment. We have
considered different classes of Gaussian states and found
the pairs of states maximizing the backflow of information.
For coherent states we have found analytical solutions,
whereas for squeezed states we have resorted to
numerical maximization, and also obtained some
approximate analytical solutions in the weak coupling limit.
\par
Our results are encouraging enough to suggest the use of our measure
of non-Markovianity to analyze more general Gaussian channels,
and to assess non-Markovianity as a resource for quantum technologies.
\section*{Acknowledgments}
This work has been supported by the Finnish Cultural Foundation
(Science Workshop on Entanglement), the Emil Aaltonen Foundation,
the Magnus Ehrnrooth Foundation, and the German Academic Exchange
Service. RV, SM and MGAP thank Stefano Olivares for useful
discussions.


\begin{thebibliography}{99}
\bibitem{BrePet}
H. P. Breuer and F. Petruccione, \emph{The Theory of Open Quantum
Systems} (Oxford University Press, Oxford, 2002).
\bibitem{Weiss}
U. Weiss, \emph{Quantum Dissipative Systems} (3rd Edition),
(World Scientific, Singapore, 2008).
\bibitem{LiGoKoSu}
G. Lindblad, Commn. Math. Phys., \textbf{48}, 119-130 (1976); V.
Gorini, A. Kossakowski, and E.C.G. Sudarshan, J. Math. Phys.
\textbf{17}, 821-825 (1976).
\bibitem{Bre09}
H.-P. Breuer, E.-M. Laine, and J. Piilo,
Phys. Rev. Lett. \textbf{103}, 210401 (2009);
E.-M. Laine, J. Piilo, and H.-P. Breuer,
Phys. Rev. A \textbf{81}, 062115 (2010).
\bibitem{NMQJ} J. Piilo, S. Maniscalco, K. H\"ark\"onen, and K.-A. Suominen
Phys. Rev. Lett. {\bf 100}, 180402 (2008);
J. Piilo, K. H\"ark\"onen, S. Maniscalco, and K.-A. Suominen, Phys.
Rev. A {\bf 79}, 062112 (2009).
\bibitem{pseudo}
L. Mazzola, S. Maniscalco, J. Piilo, K.-A. Suominen, and B. M.
Garraway Phys. Rev. A {\bf 80}, 012104 (2009).
\bibitem{Wolf08}
M.M. Wolf, J. Eisert, T.S. Cubitt, and J.I. Cirac, Phys. Rev. Lett.
\textbf{101}, 150402 (2008).
\bibitem{Rivas10}
A. Rivas, S. F. Huelga, and M.B. Plenio, Phys. Rev. Lett.
\textbf{105}, 050403 (2010).
\bibitem{Chin11}
A. W. Chin, S. F. Huelga, M. B. Plenio, e-print on arXiv:1103.1219.
\bibitem{Vas11}
R. Vasile, S. Olivares, M.G.A. Paris, and S. Maniscalco, Phys. Rev.
A \textbf{83}, 042321 (2011).
\bibitem{BraunRev}
S.L. Braunstein, P. van Loock, Rev. Mod. Phys. \textbf{77}, 513
(2005).
\bibitem{Bas11}
B. Vacchini, A. Smirne, E.-M. Laine, J. Piilo, and H.-P. Breuer,
e-print quant-ph/1106.0138.
\bibitem{Znidaric11}
M. \v{Z}nidari\v{c}, C. Pineda, and I. Garc\'{\i}a-Mata, Phys. Rev.
Lett. \textbf{107}, 080404 (2011).
\bibitem{Mar04}
P. Marian, T. A. Marian, and H. Scutaru,
Phys. Rev. A {\bf 69}, 022104 (2004).
\bibitem{Gauss}
A. Ferraro, S. Olivares, and M.G.A. Paris, \emph{Gaussian States in
Quantum Information}, (Bibliopolis, Napoli, 2005).
\bibitem{Scu98}
H. Scutaru, J. Phys. A \textbf{31}, 3659 (1998).
\bibitem{Ebe10}
T. Eberle et al., Phys. Rev. Lett. 104, 251102 (2010)
\bibitem{QBMEq}
S. Maniscalco, J. Piilo, F. Intravaia, F. Petruccione, and A. Messina, Phys. Rev. A \textbf{70}, 032113 (2004).
\end{thebibliography}
\end{document}